\documentclass[epj]{svjour}
\usepackage{latexsym}
\usepackage{graphics}
\usepackage{graphicx}
\usepackage{amsmath} 
\usepackage{caption}
\usepackage{booktabs}
\usepackage{tikz}
\usepackage{array}   
\usepackage{ulem} 
\usepackage[utf8]{inputenc}

\newcommand{\markerCircleFull}{
    \tikz[baseline={([yshift=-.5ex]current bounding box.center)}] 
        \fill[red] (0,0) circle (3pt);
}
\newcommand{\markerBoxFull}{
    \tikz[baseline={([yshift=-.5ex]current bounding box.center)}] 
        \fill[blue](0,0) rectangle(0.2,0.2);}
\newcommand{\markerBoxOpen}{
    \tikz[baseline={([yshift=-.5ex]current bounding box.center)}] 
        \draw[thick,magenta] (-0.12,-0.12) -- (0.12,-0.12) 
        (0.12,-0.12) -- (0.12,0.12)
        (0.12,0.12) -- (-0.12,0.12)
        (-0.12,0.12) -- (-0.12,-0.12);}
 
\newcommand{\markerCircleOpen}{
    \tikz[baseline={([yshift=-.5ex]current bounding box.center)}] 
        \draw[thick,teal] (0,0) circle (3pt);
}

\newcommand{\markerTriangleFull}{
    \tikz[baseline={([yshift=-.5ex]current bounding box.center)}]
        \fill[magenta] (0,0.08) -- (-0.12,-0.1) -- (0.12,-0.1) -- cycle;
}
\definecolor{Oliv}{RGB}{128, 128, 0}

\newcommand{\markerTriangledown}{
    \tikz[baseline={([yshift=-.5ex]current bounding box.center)}]
        \fill[Oliv] (0,-0.08) -- (-0.12,0.1) -- (0.12,0.1) -- cycle;
}

\newcommand{\markerTriangleOpen}{
    \tikz[baseline={([yshift=-.5ex]current bounding box.center)}]
        \draw[thick,green!70!black] (0,0.08) -- (-0.12,-0.1) -- (0.12,-0.1) -- cycle;
}

\newcommand{\markerDiamondOpen}{
    \tikz[baseline={([yshift=-.5ex]current bounding box.center)}]
        \draw[thick,blue] (0,0.14) -- (0.11,0) -- (0,-0.14) -- (-0.11,0) -- cycle;
}

\newcommand{\squareCross}{
  \tikz[baseline={([yshift=-.5ex]current bounding box.center)}]{
  
   \draw[thick,red] (-0.05,-0.15) -- (0.05,-0.15);
   \draw[thick,red] (-0.05,-0.15) -- (-0.05,-0.15);
   \draw[thick,red] (-0.05,-0.15) -- (-0.05,-0.05);
   \draw[thick,red] (0.05,-0.15) -- (0.05,-0.05);
   \draw[thick,red] (-0.05,-0.05) -- (-0.15,-0.05);
   \draw[thick,red] (-0.15,-0.05) -- (-0.15,0.05);
   \draw[thick,red] (-0.15,0.05) -- (-0.05,0.05);
   
   \draw[thick,red] (-0.05,0.05) -- (-0.05,0.15);
   \draw[thick,red] (-0.05,0.15) -- (0.05,0.15);
   \draw[thick,red] (0.05,0.15) -- (0.05,0.05);
   \draw[thick,red] (0.05,0.05) -- (0.15,0.05);
   \draw[thick,red] (0.15,0.05) -- (0.15,-0.05);
   \draw[thick,red] (0.15,-0.05) -- (0.05,-0.05);
    
  }
}
\begin{document}
%
\title{Proposal for the first measurement of antiproton polarization in proton-nucleus interactions}

\author{D.~Alfs\inst{2} \and 
        D.~Grzonka\inst{1,2}\and 
        G.~Khatri\inst{6} \and 
        P.~Kulessa\inst{4} \and 
        J.~Ritman\inst{1,3,2} \and 
        T.~Sefzick\inst{1} \and 
        J.~Smyrski\inst{5} \and 
        V.~Verhoeven\inst{3,1} \and 
        H.~Xu\inst{1} \and
        M.~Zieliński\inst{5,}\thanks{email: {marcin.zielinski@uj.edu.pl} (corresponding author)}
}

\institute{GSI Helmholtz Centre for Heavy Ion Research GmbH, 64291 Darmstadt, Germany \and 
           Institute of Nuclear Physics, Forschungszentrum J\"{u}lich, 52428 J\"{u}lich, Germany \and 
           Ruhr-University Bochum, 44801 Bochum, Germany \and 
           H.~Niewodniczański Institute of Physics, Polish Academy of Science, 31-342 Kraków, Poland \and 
           M.~Smoluchowski Institute of Physics, Jagiellonian Univeristy, 30-348 Kraków, Poland \and 
           CERN European Organization for Nuclear Research, 1211 Geneva 23, Switzerland
}

\date{Received: date / Revised version: date}
%
\abstract{Spin dependent phenomena in inclusive hadron production have been extensively investigated, yet their microscopic origin and universality across different hadrons are still not fully understood. In particular, it is presently unknown whether antiprotons produced in unpolarized hadronic collisions can acquire a transverse polarization as a result of spin dependent $\bar{p}N$ interactions and nonperturbative hadronization mechanisms. Establishing the presence or absence of such an effect would provide new empirical constraints on the spin structure of the antinucleon-nucleon interaction, which is only weakly constrained by existing data.
In this work, we investigate the experimental feasibility of a first dedicated measurement of the transverse polarization of antiprotons produced in proton-nucleus collisions. The polarization is accessed through the left-right asymmetry in elastic $\bar{p}p$ scattering in the Coulomb Nuclear Interference region. Based on detailed Monte Carlo simulations of the proposed experimental setup at the European Organization for Nuclear Research (CERN), we estimate the statistical sensitivity required to detect a certain degree of polarization.
} 
\maketitle

\section{Introduction}
\label{sec:intro}
The existence of antiprotons was discovered in 1955 with the Bevatron~\cite{Chamberlain:1955ns} and the inaugural containment of antiprotons was first achieved in 1978 by the Initial Cooling Experiment~\cite{Krienen1980} conducted at CERN. 
Since then, antiprotons have become a versatile tool in particle physics, with beams being routinely produced at several facilities~\cite{Caravita:2025peq,Ponce:2022jbb} and future projected~\cite{Spiller:2006gj}. They have enabled precision studies of fundamental symmetries, antimatter properties, and nucleon–antinucleon interactions across a broad range of energies~\cite{ALPHA:2018sre,ATRAP:2012rpd}. 
Despite this progress, the question of how antiprotons can be efficiently polarized, a prerequisite for a new class of experiments, remains open. Moreover, this constitutes a significant gap in our understanding of spin phenomena in the baryon–antibaryon sector.

Although significant experimental datasets and multiple theoretical approaches have emerged in recent years, a satisfactory description of spin dependent effects in inclusive hadron production, including antiprotons, within QCD has not yet been achieved. Whereas polarized proton beams have long been exploited to study spin observables, the possibility that antiprotons may exhibit polarization, generated directly by strong interaction dynamics rather than external spin manipulation, has never been experimentally investigated.

Unlike the case of antiprotons, the polarization of hyperons has been observed in a wide variety of high energy $pp$ and $pN$ experiments~\cite{Bunce:1976yb,Heller:1978ty,Erhan:1979xm,Gourlay:1986mf,Lundberg:1989hw,RAMBERG1994403,HERA-B:2006rds,ATLAS:2014ona,HADES:2014ttv}. 
These measurements show that sizable transverse polarization, can emerge from nonperturbative QCD dynamics even in the absence of initial state polarization~\cite{Troshin:2002xs}. This is especially well seen for $\Lambda$ hyperons, which show a significant degree of polarization when produced in collisions of high energy unpolarized protons~\cite{RAMBERG1994403}. 
If an analogous effect exists for antiprotons, its observation would provide direct empirical access to spin–orbit components and spin–flip amplitudes of the $\bar{p}N$ interaction elements that are poorly constrained due to the absence of dedicated measurements. The description of the spin dependent structure of the antinucleon-nucleon ($\bar{N}N$) is important for understanding the fundaments of the strong interaction. While the long range part of the $\bar{N}N$ force can be related to the nucleon-nucleon system through meson exchange models and $G$-parity arguments, the short-range dynamics of $\bar{N}N$ interactions differ qualitatively due to the presence of strong annihilation channels already at low relative momenta\cite{Richard:2019dic,Richard:2022tpn}. Therefore, establishing whether antiprotons emerging from $pp(A)$ reactions carry measurable transverse polarization~\cite{Kilian:2011zz} is not only a prerequisite for future polarized antiproton beams, but also a genuinely new test of strong interaction dynamics in the baryon–antibaryon sector.

Currently, no efficient method exists for producing a suitably polarized antiproton beam for experiments without significant modifications to the existing and planned infrastructure. The first collection of ideas to produce polarized antiproton beams was presented in 1985~\cite{10.1063/1.35683} and more recently described in~\cite{1996PhRvL..77.2626B,2009AIPC.1149...80S,2008AIPC.1008..124M,2010EPJA...43....5S}.
The proposed mechanisms cover a large field of methods, such as hyperon decay, spin filtering, spin flip processes, stochastic techniques, dynamic nuclear polarization, spontaneous synchrotron radiation, induced synchrotron radiation, interaction with polarized photons, Stern-Gerlach effect, channeling, and polarization of trapped antiprotons, antihydrogen atoms, as well as the initial polarization of produced antiprotons. 
Some of these methods have already been discarded due to the low intensity or polarization of the resulting antiproton beam. 

Therefore, the aim of this article is to determine the conditions under which the transverse polarization of produced antiprotons may be observable. Using precise Monte Carlo calculations, the necessary conditions needed to observe a specified degree of polarization is estimated.
In the case where a non-zero initial polarization is present, the preparation of a polarized antiproton beam can be achieved with relatively modest additional effort. If the produced antiprotons exhibit sufficient polarization, a selection of specific azimuthal regions in the production line would yield a polarized secondary beam directly. The feasibility studies presented in this work form the basis for an experimental investigation within the P371 experiment at the CERN T11 beam line.

\section{Polarization in hyperon production and its implications for antiprotons}

The spin dependent observables have historically played a central role in testing the strong interaction dynamics. Sizable hyperon polarizations, observed over a broad range of energies, remain among the phenomena for which QCD does not provide a unique or quantitatively complete explanation. Existing experimental datasets reveal a highly nontrivial phenomenology. The first observation of polarization in unpolarized $pp$ and $pA$ collisions was reported for $\Lambda$ hyperons at Fermilab~\cite{Bunce:1976yb}, where unexpectedly large transverse polarization in the order of $20$--$30\%$ in the forward region was measured despite the unpolarized proton beams. Subsequent dedicated studies of hyperons~\cite{Heller:1978ty,RAMBERG1994403} confirmed that transverse polarization persists over a broad energy range and follows a characteristic dependence on the transverse momentum and $x_{F}$ of the produced hyperons.

This persistent observation of large polarization has stimulated extensive theoretical activity aimed at identifying its microscopic origin. A variety of phenomenological approaches have been developed to describe hyperon polarization and related analyzing power data~\cite{Liang:1997rt,Boros:1998kc,Troshin:1995rt,Troshin:1996hd,DeGrand:1981pe,Degrand:1985xm,Andersson:1979wj}. These models emphasize different aspects of nonperturbative QCD dynamics, including the role of the orbital angular momentum of valence quarks, parton rotation inside constituent quarks, spin-dependent recombination mechanisms, and polarized string fragmentation. While these approaches differ in their underlying assumptions and degrees of phenomenological input, they share the common feature that transverse polarization emerges from the interference of spin dependent amplitudes during the hadron formation process.

In semiclassical approaches based on valence quark dynamics, transverse hyperon polarization is attributed to spin–orbit correlations arising from the orbital motion of valence quarks during hadron formation~\cite{Liang:1997rt,Boros:1998kc}. The polarization magnitude and its characteristic $x_F$ dependence are linked to the number of valence quarks shared between the produced hyperon and the incident hadron. Within this framework, qualitative agreement is obtained for the sign and relative strength of $\Lambda$, $\Sigma^{-}$, and $\Xi^{0,-}$ polarizations in $pp$ collisions, as well as for the reduced polarization observed with pion beams. However, these models do not provide a quantitative description of antihyperon polarization and leave the origin of the required phase differences largely unconstrained.

Another model attributes hyperon polarization to the internal dynamics of constituent quarks, particularly to the rotation of quark-antiquark pairs inside their extended structure~\cite{Troshin:1995rt,Troshin:1996hd}. In this picture, the polarization of strange quarks arises from multiple scattering of constituent quarks in an effective mean field, leading to correlations between spin and internal orbital angular momentum. The model predicts a negative polarization of $\Lambda$ hyperons with a characteristic dependence on $x_F$ and scaling behavior at sufficiently large transverse momenta. While this framework provides a qualitative description of $\Lambda$ polarization systematics, it does not offer quantitative predictions for other hyperons and does not explicitly constrain the phase structure required for a complete description of transverse polarization phenomena.

In recombination based approaches, transverse hyperon polarization is attributed to spin–momentum correlations generated during the recombination of partons into the final state hadron~\cite{DeGrand:1981pe,Degrand:1985xm}. A key ingredient of these models is the Thomas precession~\cite{Thomas:1927yu} of quark spins, induced when the force acting on a quark is not aligned with its velocity during the hadronization process. This mechanism naturally leads to a characteristic dependence of the polarization on transverse momentum and $x_F$, and it reproduces the observed sign of $\Lambda$ polarization as well as its approximate scaling behavior. While the model captures several  features of the data, it does not provide a consistent quantitative description across different hyperon species and reaction channels, and it predicts vanishing polarization for antibaryons, in contrast to the complex experimental phenomenology. The strong sensitivity of the predicted polarization to assumptions about the hadronization time and effective quark masses further indicates that recombination models probe nonperturbative aspects of confinement dynamics.

Within the Lund string fragmentation framework, transverse hyperon polarization is attributed to spin–momentum correlations generated during the breaking of a color flux tube stretched between the outgoing parton and the remnant hadronic system~\cite{Andersson:1979wj}. Assuming SU(6) wave functions for hadrons, the polarization of the $\Lambda$ hyperon is identified with the polarization of the strange quark produced during string breaking. In this picture, local transverse momentum compensation between the $s$ and $\bar{s}$ quarks leads to an orbital angular momentum of the pair, which is correlated with their spin orientation and results in a non-zero polarization normal to the production plane. The model correctly predicts the sign of $\Lambda$ polarization in $pp$ collisions and a linear dependence on transverse momentum, but it does not provide quantitative predictions for the polarization magnitude or its $x_F$ dependence. Moreover, additional assumptions are required to extend the description to other reactions, and the model does not account for antihyperon polarization.

Despite their conceptual differences, all phenomenological approaches discussed above emphasize that transverse hyperon polarization originates from nonperturbative spin dependent dynamics and from the interference of amplitudes with nontrivial phase structure. This common feature provides a natural motivation to explore whether analogous mechanisms can generate polarization for produced antiprotons. To date, no measurement of polarization for directly produced antiprotons has been reported, either in $pp$, $pA$, or heavy ion collisions. As a result, the possibility of non-zero antiproton polarization constitutes an open experimental question.

\section{Motivation for the use of polarized antiprotons}
\label{sec:motivation}
One of the most attractive uses of polarization observables is to obtain a more detailed insight into the inner structure of hadrons and their interactions, with the deciphering of diverse reaction mechanisms often hinging on the control of spin orientations.
By polarizing both beam and target particles, one can selectively populate quantum states. Therefore, access to  polarized antiproton beams would unlock a multitude of experiments across the spectrum of low to high energy physics~\cite{KLEMPT2002119}. The ability to manipulate various spin configurations holds significance across a spectrum of topics. 

Among the most fundamental problems accessible with polarized antiprotons is the characterization of the antinucleon–nucleon ($\bar{N}N$) interaction~\cite{PHILLIPS:1967wls,Bystricky1978,Dover:1992vj,Richard:2022tpn}.
For instance, in $\bar{p} - p$ reactions, a parallel spin alignment of the antiproton and proton ($\bar{p}^\uparrow$ $p^\uparrow$) forms a specific spin triplet state, whereas an anti-parallel spin alignment ($\bar{p}^\uparrow$ $p^\downarrow$) predominantly yields a spin singlet state. The long-range $\bar{N}N$ interaction is described through meson exchange models and can be correlated with the $NN$ system via the G-parity rule. However, the short-range interaction diverges fundamentally between these two systems: while the inelastic component of the $NN$ cross section undergoes a significant surge for beam momenta exceeding approximately 1~GeV/c, the inelastic channels for the $\bar{N}N$ system manifest even at the lowest beam momenta.

Currently, only a handful of spin dependent observables have been investigated for $\bar{p}p$ scattering. These include measurements conducted at the Low Energy Antiproton Ring (LEAR)~\cite{KLEMPT2002119}, which yielded insights into the $\bar{N}N$ interactions, and experiments at FERMILAB~\cite{FNAL-E581704:1989hno,E581704:1996bnv}, where polarized antiproton beams were employed to study the spin dependence of the total $\bar{p}p$ cross section for longitudinally polarized states. However, there are no experimental data that offer comprehensive insights into the spin dependent aspects of the antiproton-proton scattering cross section $\sigma_{\text{tot}}$ given by~\cite{Bystricky:1976jr}:
\begin{equation}
\sigma_{\text{tot}} = \sigma_{0} + \sigma_{1} \left( \overset{\rightarrow}{P} \cdot \overset{\rightarrow}{Q} \right) + \sigma_{2} \left( \left(\overset{\rightarrow}{P}\cdot \hat{k}\right) \left(\overset{\rightarrow}{Q} \cdot \hat{k}\right) \right), 
\end{equation}
where $\overset{\rightarrow}{P}$ and $\overset{\rightarrow}{Q} $ are beam and target polarizations, $\hat{k}$ is the direction along the beam momentum, $\sigma_{0}$ is the unpolarized scattering cross section, and $\sigma_{1}$ and $\sigma_{2}$ are the spin-dependent cross sections for transverse and longitudinal spin, respectively.
In the previous experiments, only $\sigma_{0}$ was measured over a broad range of energies, while $\sigma_{1}$ and $\sigma_{2}$ remain unknown. For example, the lack of knowledge about the values of these observables makes it difficult to estimate the efficiencies for polarizing antiprotons using methods such as spin flipping or spin filtering. 

Another important topic for which antiproton polarization would be important is the study of nucleon quark structure and the origin of hadron spin.
In Quantum Chromodynamics (QCD) the hadronic quark structure~\cite{BARONE20021} requires knowledge of the unpolarized quark distribution function, the longitudinal polarization (helicity) distribution, and the transversity distribution. In particular, the transversity distribution cannot be measured directly in deep inelastic scattering experiments, nor can it be derived from other observables. This could only be achieved by directly measuring the transversity (transverse polarization) distribution in Drell-Yan lepton pair production, which requires a polarized target and beam. Such studies in a broad kinematic range were proposed at RHIC~\cite{Bunce:2000uv} and by the Polarized Antiproton eXperiment (PAX) Collaboration~\cite{10.1063/1.2122212,Lenisa:2005is}. 

\section{Techniques for polarized antiproton production}
\label{sec:methods}
Up to now, there has been only one successful demonstration of a working polarized antiproton beamline at FERMILAB~\cite{FNAL-E581704:1989hno,PhysRev.129.1795}. The source of the polarized antiprotons was the parity violating decay of the $\bar{\Lambda}$ antihyperon that decayed into an antiproton and a positively charged pion. The unpolarized $\bar{\Lambda}$'s were produced by the 800 GeV/c primary proton beam scattered on a beryllium target. In the $\bar{\Lambda}$ rest frame, the decay occurs isotropically, and the $\bar{p}$ produced are polarized at 76\% along their direction of motion~\cite{ParticleDataGroup:2024cfk}. Transforming from the rest frame to the laboratory frame introduces a transverse spin component. Selecting antiprotons that decay around $\theta = 90^\circ$ in the $\bar{\Lambda}$ reference frame results in nearly complete transverse polarization of the beam in the laboratory frame. The tagging system used allowed for the selection of $\bar{p}$ with a mean polarization of 45\% and an intensity of 10$^{4}~\bar{p}$/s, at 185~GeV/c momentum. 
This technique relies on the production of highly boosted antihyperons, which provide both a large decay length and a favorable kinematic separation of decay products. As a result, it is naturally suited to high energy environments. At lower beam momenta, antihyperon production cross sections decrease rapidly, while the reduced boost shortens the decay length and limits the phase space region in which a transverse polarization component can be efficiently selected. In principle, polarized antiprotons produced at high energies from antihyperons decay could subsequently be decelerated to lower momenta using cooling and deceleration techniques. However, the preservation of polarization during large momentum changes and long cooling times is non trivial. Depolarization effects associated with spin resonances and stochastic or electron cooling processes may lead to substantial polarization losses. In addition, the overall beam intensity would be further reduced by the cumulative efficiency of production, selection, deceleration, and cooling stages. For these reasons, while the antihyperon decay method remains, in principle, applicable beyond the originally demonstrated energy regime, its practical realization at low energies represents a significant technical challenge.

Another approach that has been studied for producing polarized $\bar{p}$ beams is the spin–filtering method.
The technique was initially proposed in 1968 at CERN for high energy protons using the Intersecting Storage Rings (ISR)~\cite{CSONKA1968247}. In this approach, an initially unpolarized beam circulates in a storage ring and interacts with a polarized target. This technique leverages the inherent spin dependence of hadronic cross sections, resulting in a differential scattering rate for particles of distinct spin projections circulating within the system. The disadvantage of this method is the reduction in beam intensity due to the systematic loss of scattered particles from the beam. To maintain the highest possible polarization, it is important to minimize the loss of particles over time. In 1992, the Test Storage Ring (TSR) experiment in Heidelberg demonstrated that an initially unpolarized proton beam can acquire polarization through spin dependent interactions with a polarized hydrogen gas target. 
In the TSR experiment, the proton beam polarization reached approximately 2\% after 90~minutes of circulation, while its intensity decreased to approximately 5\% of the original value~\cite{Rathmann:1993xf}. The study confirmed the feasibility of the spin--filtering method and measured an effective polarizing cross section of $\sigma_{\text{eff}} = 72.5$~mb.
The same approach may be used for antiprotons if a filter interaction with large spin-spin dependence and cross section is found. However, currently there are no data available on the spin-spin dependence of the total antiproton-proton cross section.
This method was under investigation by the PAX collaboration, which successfully demonstrated the feasibility of the spin-filtering method at the COSY ring for transverse polarized protons~\cite{AUGUSTYNIAK201264}. 

In addition to the previously explored polarization techniques, the spin dependent mechanisms discussed in Sec.~2 motivate the investigation of whether a non-zero polarization is generated already at the antiproton production stage. As mentioned before, in many experiments, it has been shown that the direct production of hyperons induced by high energy collisions of unpolarized protons with a solid target results in a significant degree of polarization~\cite{Heller:1978ty,Bunce:1976yb,R608:1986ltk,Ho:1990dd,E761:1993qya,RAMBERG1994403}. 
Despite the different production mechanisms of antiprotons and hyperons, in which a strange quark needs to be created, it is probable that the production process itself can be a source of particle polarization. 
For antiprotons produced in a strong process via the reaction $p+p \to pp\bar{p}p$, a potential rise in polarization could be a manifestation of the spin-orbit interaction. 
The minimum antiproton transverse momentum in the laboratory frame needed to include more than pure S-wave is about 197 MeV/c (roughly estimated by the angular momentum $L = r \times p$ at about 1~fm distance).
If transverse polarization occurs, then a polarized beam can be prepared by selecting events emitted in a narrow azimuthal range. Furthermore, the pure S-wave region near polar angles of 0 degrees ($\leq$ 50 mrad) has to be removed. 

Up to now, a possible polarization of produced antiprotons has never been measured experimentally.
Therefore, in the P371 experiment, this approach will be explored in view of delivering beams of polarized antiprotons and studying $\bar{N}N$ strong interaction dynamics.

\section{Polarization determination}
\label{sec:polarozation}
To observe the possible transverse polarization of produced antiprotons, the asymmetry needs to be determined in a reaction with a known analyzing power. The cross section for the scattering of transversely polarized antiprotons on an unpolarized target is given by: 
\begin{equation}
    \frac{d\sigma}{(d\theta d\phi)} = \frac{d\sigma_{0}}{(d\theta d\phi)}\left( 1 + A_{y}(\theta) P \cos\phi\right),
\end{equation}
where $\sigma_{0}$ is the cross section for an unpolarized beam, $A_{y}$ is the single-spin asymmetry (the analyzing power), $P$ is the polarization, and $\theta$ and $\phi$ are the polar and azimuthal angles, respectively. Hence, it is important to choose a kinematic region with a well known and sufficiently large analyzing power $A_{y}$. Experimentally, the polarization can be calculated by determining the asymmetry in the number of antiprotons scattered to the left and to the right side, measured in the coordinate system given by the $\phi$ angle between the positive side of the y-axis and the spin projection onto the xy-plane. In this case, the asymmetry $ \epsilon $ can be written as:
\begin{equation}
    \epsilon = \frac{N_L - N_R}{N_L + N_R} = A_y (\theta) P \cos\phi,
\label{eq:asymmetry}
\end{equation}
where $N_L$ and $N_R$ are the number of antiprotons scattered to the left and right side, respectively. 

A well suited reaction for polarization studies is the elastic antiproton-proton scattering.
The elastic interaction between charged hadrons is a mixture of the strong and electromagnetic interactions, where Coulomb interactions are dominant for small four-momentum and the strong interaction dominates in the large momentum transfer region.
In this view, the most convenient region for a polarization measurement is the Coulomb-Nuclear Interference (CNI) region, where the nuclear and Coulomb interactions are of similar strength and for which $A_{y}$ is well known and calculable for the proton-proton system. Therefore, for the CNI region, the unpolarized differential cross section is given by the sum of hadronic ($had$), electromagnetic ($em$), and interference ($int$) contributions~\cite{E581704:1993mgc}:
\begin{equation}
    A_{y}\frac{d\sigma_0}{dt} = \left(A_{y}\frac{d\sigma_0}{dt}\right)^{had} 
                              + \left(A_{y}\frac{d\sigma_0}{dt}\right)^{em} 
                              + \left(A_{y}\frac{d\sigma_0}{dt}\right)^{int}. 
\end{equation}
\begin{figure}[t]
\centering
\resizebox{0.425\textwidth}{!}{%
  \includegraphics{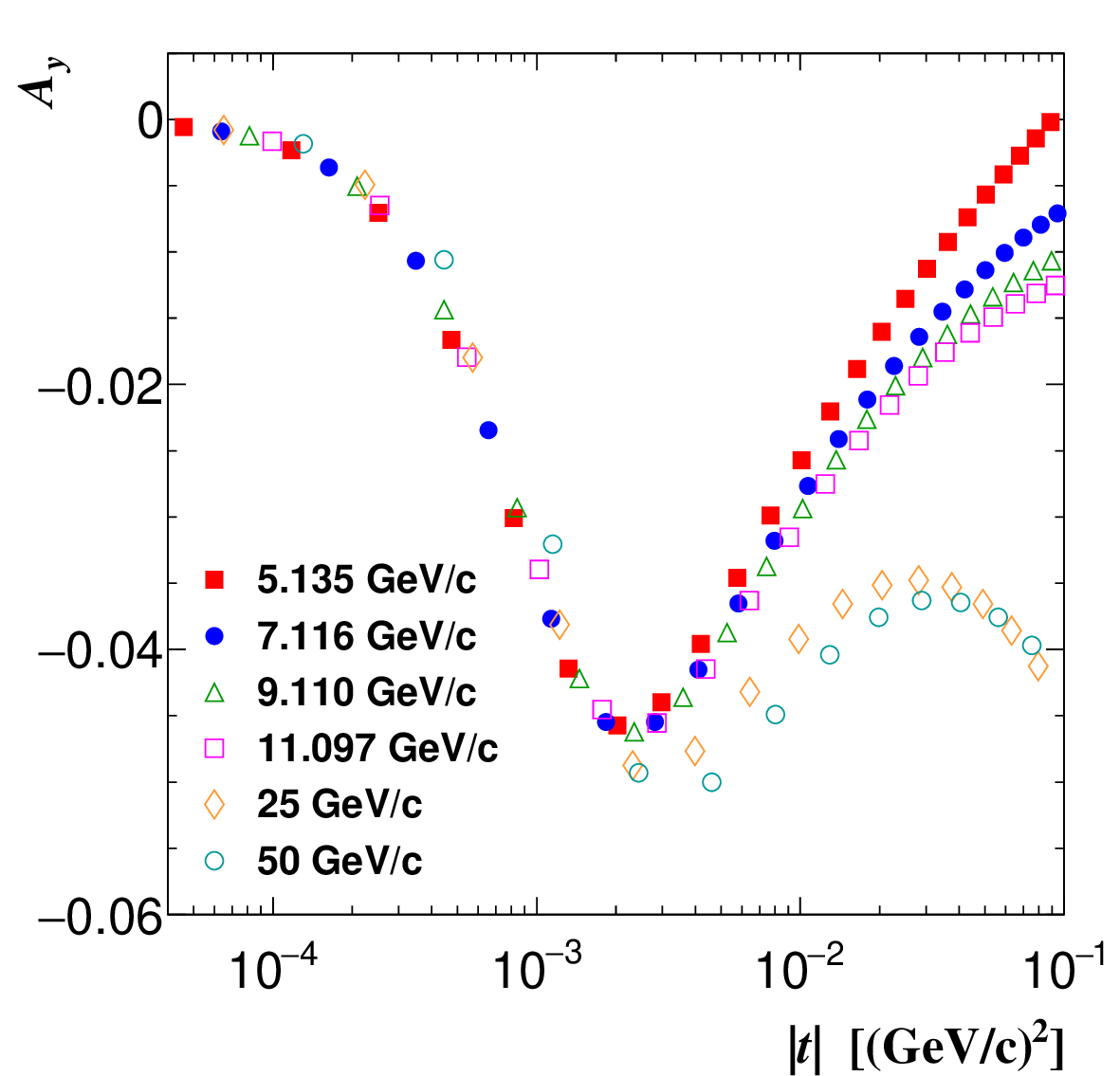}
}
\caption{\label{fig:haidenbauer} Preliminary calculations of $A_y$ within the one-boson exchange model based on $\bar{N}N$ potentials adjusted to experimental data, performed by J.~Haidenbauer~\cite{Haidenbauer:2014rva,Haidenbauer:2017crp,Haidenbauer-pc}. 
Points colors represent $A_y$ values obtained based on the model calculations for existing experimental data on $\bar{p}p$ scattering measured for different momenta as indicated in the figure. 
}
\label{fig:1}       
\end{figure}
In this approach, the leading terms of the electromagnetic contributions are described by the one-photon exchange approximation known from Quantum Electrodynamics (QED), resulting in $A_{y}^{em} \sim  0$. Moreover, at high energies and low four-momentum transfer, $A_{y}^{had}$ is also expected to vanish~\cite{E581704:1993mgc}.
Therefore, in the CNI region at high energies, $A_{y}$ is attributed to the interference between a non-spin-flip nuclear amplitude and an electromagnetic spin-flip amplitude. For such a case, the maximum analyzing power can be approximated by~\cite{Kopeliovich:1974ee,E581704:1993mgc}:
\begin{equation}
    A_{y}^{max} = \frac{\sqrt{3}}{4}\frac{t_p}{m}\frac{(\mu - 1)}{2},
\end{equation}
where $m$ is the proton mass, $\mu$ denotes the proton magnetic moment, and $t_p$ is the four-momentum transfer equal to 
$t_p = -8\pi\sqrt{3}\alpha/\sigma_{tot} \approx -3\times 10^{-3}~($GeV/c$)^{2}$,
with the fine structure constant $\alpha$ and the total cross section $\sigma_{tot} = 40$~mb~\cite{ParticleDataGroup:2024cfk}. These calculations were confirmed in experiments at Fermilab~\cite{E581704:1993mgc} and RHIC~\cite{OKADA2006450} using 200 GeV/c and 100~GeV/c polarized proton beams scattered on an unpolarized target, respectively, yielding a maximum analyzing power $A_{y} = 4.5\%$ for $t$ = 0.003~(GeV/c$)^{2}$. Moreover, for antiprotons, a similar analyzing power is anticipated due to G-parity conservation, but with a negative sign because only the electromagnetic amplitude reverses. This was also experimentally confirmed at Fermilab with a 185 GeV/c antiproton beam that resulted in $A_{y} = (-4.6 \pm 1.86)\%$~\cite{E581704:1989vcm}.
However, such assumptions do not necessarily hold at lower energies because the CNI region shifts, and the cross section for proton-antiproton scattering changes significantly in comparison to the high energy case. The experimentally observed total cross section at low energies is about a factor of 2 higher than 40~mb at high energy~\cite{ParticleDataGroup:2024cfk}. 

Therefore, for the beam momenta of 3.5~GeV/c, the assumed analyzing power has to be anticipated based on theoretical predictions. Preliminary calculations were performed within a one boson exchange model, using an antinucleon–nucleon potential adjusted to existing experimental data on $\bar{p}p$ scattering for beam momenta ranging from 50~GeV/c down to 5~GeV/c~\cite{Haidenbauer:2014rva,Haidenbauer:2017crp,Haidenbauer-pc}.
The resulting predictions for the analyzing power $A_{y}$  as a function of four momentum transfer $t$ for different beam momenta are presented in Fig.~\ref{fig:haidenbauer}. The $A_y$ distributions at lower momenta are found to be similar to those at higher energies, exhibiting comparable amplitudes of about $4.5\%$ at smaller values of $t$ (around $|t_p| = 0.0025~\mathrm{GeV}/c^{2}$). Moreover, the shape of the function remains largely unchanged for beam momenta at or below 11~GeV/c. Based on these predictions, for momenta 3.5~GeV/c in the selected CNI region, which corresponds to an angular range of up to 20~mrad, the expected analyzing power should be approximately $4.5\%$.

\section{Feasibility studies}
\begin{figure*}[ht]
\centering{
\resizebox{0.90\textwidth}{!}{%
  \includegraphics{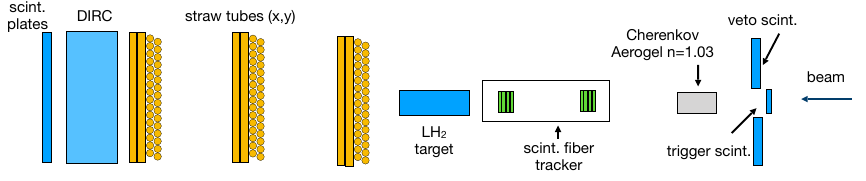}
}}
\caption{Scheme of the proposed experimental setup for determining the antiproton polarization. The beam enters from the right and passes  through a scintillator detector (start/trigger), Cherenkov detectors, fiber detectors, the analyzing target (LH$_2$), straw tube detectors, a DIRC counter, and scintillator detectors (stop/trigger).}
\label{fig:detectors}      
\end{figure*}
\label{sec:studies}
At proton momenta of a few GeV/c, antiproton production in proton–nucleus collisions seems to proceed predominantly via the quasi–free $p+p \to pp\bar{p}p$ channel. However, multi step nuclear processes, pion induced reactions, and the production of neutrons contribute significantly and could dominate the yield. The threshold momentum for the antiproton production
in a $pp$ collision is about 6~GeV/c, and the antiprotons are produced isotropically in the center of mass of the colliding nucleons. However, to achieve a useful yield of antiprotons, higher energies of the primary proton beam are used, in CERN it is 24~GeV/c momentum, which results in a broad antiproton momentum distribution with an intensity maximum around 3.5~GeV/c~\cite{Eichten:1972nw}.
To obtain non-zero polarization of antiprotons, a necessary condition is that the production polar angle is large enough to avoid the pure S-wave region ($\leq 50$~mrad). 

\subsection{Experimental setup design and beamline conditions}\label{subsec:setup}
To study the polarization of the antiprotons, the azimuthal $\phi$ angular distribution of their elastic scattering must be precisely determined. The proposed studies are foreseen at the T11 PS CERN beamline, which can transport secondary positively and negatively charged particles with momenta of 3.5~GeV/c~\cite{Bernhard:2018ami,Bernhard:2021jeb}. 
The T11 beamline typically provides negatively charged particles with a flux of $1\cdot 10^6$ particles per 400~ms cycle, with a momentum spread of about $\pm5\%$.
However, there is no separation of antiprotons from other secondary produced particles, such as negative pions. Therefore, the desired detection apparatus should include a tracking system for the primary and scattered particles, along with particle identification detectors.
The proposed experimental setup consists of scintillators to provide the trigger signal, a threshold aerogel Cherenkov detector for pion discrimination at the trigger level, a scintillating fiber hodoscope to measure the tracks of produced antiprotons, an analyzer target, a set of straw tube detectors to reconstruct the tracks of the scattered antiprotons, and a DIRC detector for particle identification. The proposed detector scheme is presented in Fig.~\ref{fig:detectors}.  

The analyzer target is a 120 mm long and 60 mm in diameter cylindrical liquid hydrogen cell (LH\(_2\)) with 50 $\mu$m Kapton foils, placed in a vacuum chamber with windows also made of 75 $\mu$m Kapton foils, operated at a temperature of approximately 19 K.
The scintillating fiber detector consists of two stations separated by 0.4~m, each with two overlapping layers with 0.5 mm thick square fibers.
The straw tube detectors, with a tube diameter of 10~mm, are placed in three positions along the beam line and achieve position resolution 
in the range of 100 - 150~\(\mu\)m~\cite{Smyrski:2018snv,Firlej:2023paa,Mertens:2014dxa}.
The aerogel Cherenkov detector, with a refractive index of n \(\approx\) 1.03, is used to discriminate the high pion background by including its signals in the trigger system as a veto.
Antiprotons with 3.5 GeV/c momentum have a velocity of $\beta = 0.966$, which corresponds to the threshold for Cherenkov light emission of n = 1.035. 
Pions have a velocity close to the speed of light ($\beta = 0.999$), which corresponds to a threshold refractive index of n = 1.001.
Furthermore, the DIRC detector equipped with Plexiglas as a radiator and a photomultiplier matrix will be used for offline particle identification ($\pi^-/\bar{p}$)~\cite{Zink:2014qra}. Finally, behind the DIRC detector is a wall consisting of four 20 cm wide scintillating paddles, which are read out at both ends by photomultipliers. 
This acts as a stop detector, ensuring a consistent track throughout the entire setup in coincidence with the start detector. Details of the detection system and its performance will be provided in a separate publication.

\subsection{Monte Carlo event sample preparation}\label{subsec:MC}
In order to measure the degree of antiproton polarization, the asymmetry must be determined with sufficient precision.
The corresponding Monte Carlo (MC) studies were performed using the GEANT 4 simulation framework~\cite{GEANT4:2002zbu,Allison:2016lfl}.
The beam of unpolarized antiprotons was defined with the parameters specific to the T11 beam line at a nominal momentum of $p = 3.5$~GeV/c.
\begin{figure*}[!h]
\centering{
\resizebox{0.80\textwidth}{!}{%
 \includegraphics[width=0.425\columnwidth]{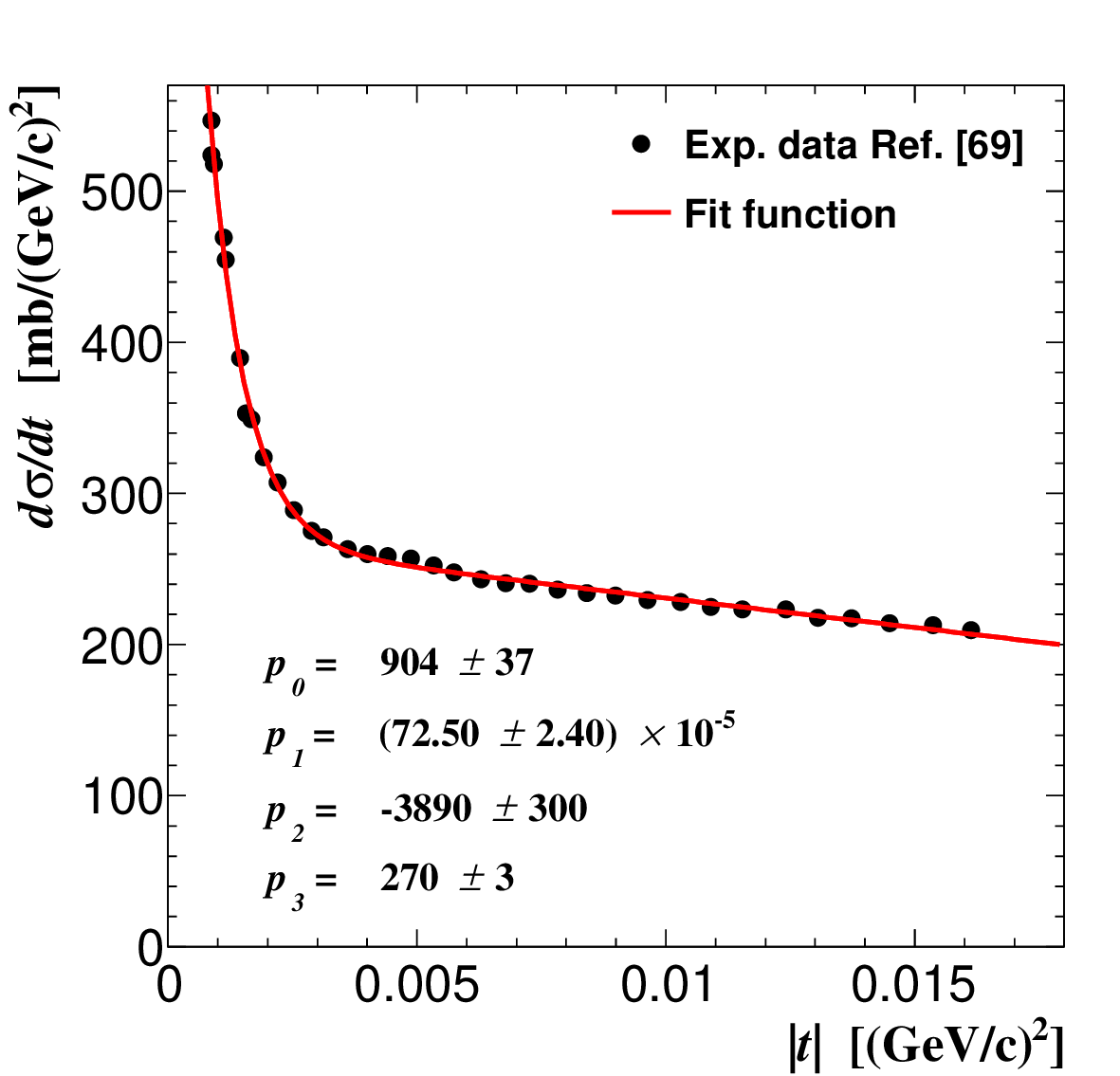}%
 \includegraphics[width=0.425\columnwidth]{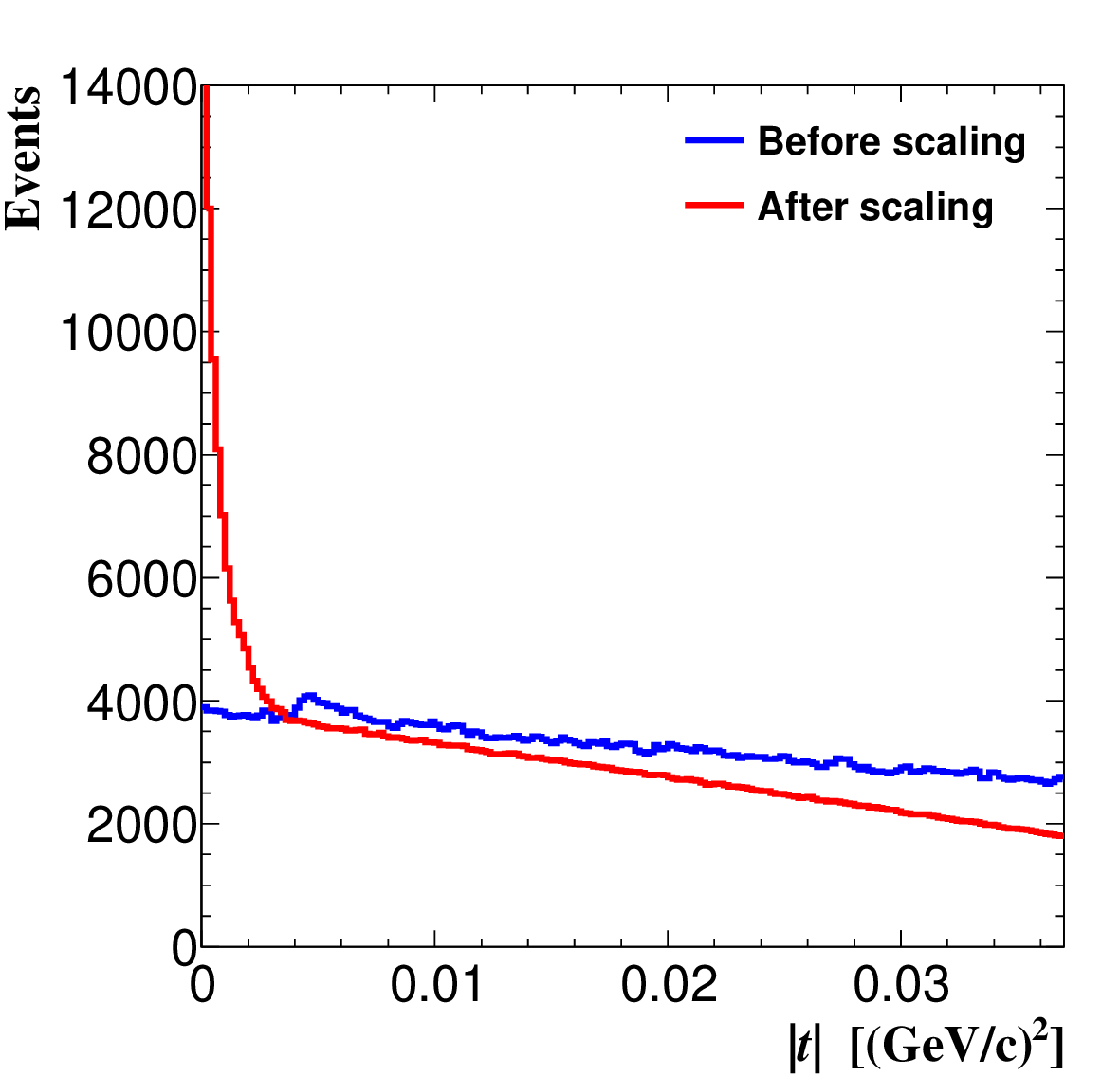}%
}}
\caption{{\bf (left)} Measured differential cross section of $p\bar{p}$ forward elastic scattering for the beam momentum $p_{\text{lab}}~=~3.7$~GeV/c~\cite{E760:1996mur} (black points). The superimposed red line indicates the fit function ${d\sigma}/{dt} = p_0(\exp({-t}/{p_1}) + p_{2}t + p_3)$. 
{\bf (right)} Number of events as a function of the four-momentum transfer where blue histogram denotes events before scaling and superimposed red histogram represents events after scaling using the weight $\omega_s(\vert t \vert)$ factor. }
\label{fig:cross-section-t}
\end{figure*}
The direction of the beam was defined to be the $z$-axis, with its mean position passing through the center of the target, and a Gaussian position distribution of $\sigma_{x} = 10$~mm, $\sigma_{y} = 5$~mm, and an angular divergence of $\sigma_{x^{\prime}} = 4$~mrad and $\sigma_{y^{\prime}} = 10$~mrad. The size and position of the beam were defined at the focal point located at the entrance of the (LH\(_2\)) target. In the experiment, most of the beam particles pass through the setup without elastic scattering within the target volume. Therefore, only events in which elastic scattering occurred within the target volume were selected for further analysis, for which the four-momentum transfer was then determined from the exact values of the antiproton four-momentum before and after the target.
For the models implemented in the simulation software, the elastically scattered events follow the expected differential cross section over a broad range of four-momentum transfer. However, at small four-momentum transfer, they deviate from the experimental data because of the lower cutoff used in the calculation of Coulomb scattering in the applied model.
Therefore, to account for this effect in the simulated sample, each event in the four-momentum transfer range $0 \leq |t| \leq 0.04$~(GeV/c)$^2$ was weighted by a scaling factor $\omega_s(|t|)$. The scaling factor is derived for each $|t|$ from the ratio of the parametrization of the measured differential cross section and the distribution of the simulation.
For this purpose, existing data~\cite{E760:1996mur} for the differential cross section of forward elastic scattering of $\bar{p}p$ at the beam momentum $p_{\text{lab}}$ = 3.7~GeV/c is used under the assumption that they are also comparable for $p_{\text{lab}}$ = 3.5~GeV/c.
Figure~\ref{fig:cross-section-t} shows the fit result for $p_{\text{lab}}$ = 3.7~GeV/c data (left) and the Monte Carlo sample before and after scaling (right). This assumption is based on previous measurements of the differential cross section of $\bar{p}p$ forward elastic scattering for six different beam momenta given in~\cite{E760:1996mur}. 
Furthermore, to implement the detector resolution, the tracks in azimuthal and polar angles were smeared by a Gaussian distribution of $\sigma = 1$~mrad. 
In order to introduce an asymmetry in the scattering process, each event was weighted by another factor $\omega_p(|t|,\phi) = 1 + P\cdot A_y(|t|)\cos\phi\sin\beta$, where $P$ is the polarization, $A_y$ is the analyzing power, and $\phi$ is the angle between the positive side of the $y$-axis and the spin projection onto the $xy$-plane, while $\beta$ is the angle between the $z$-axis and the vector of the $\hat{S}$ spin direction. With this procedure, the event sample could be generated with an asymmetry resulting from any chosen polarization. 

\subsection{Analysis of the Monte Carlo data sample}\label{subsec:MCanalysis}
\begin{figure*}
\centering{
\resizebox{0.80\textwidth}{!}{%
    \includegraphics[width=0.425\columnwidth]{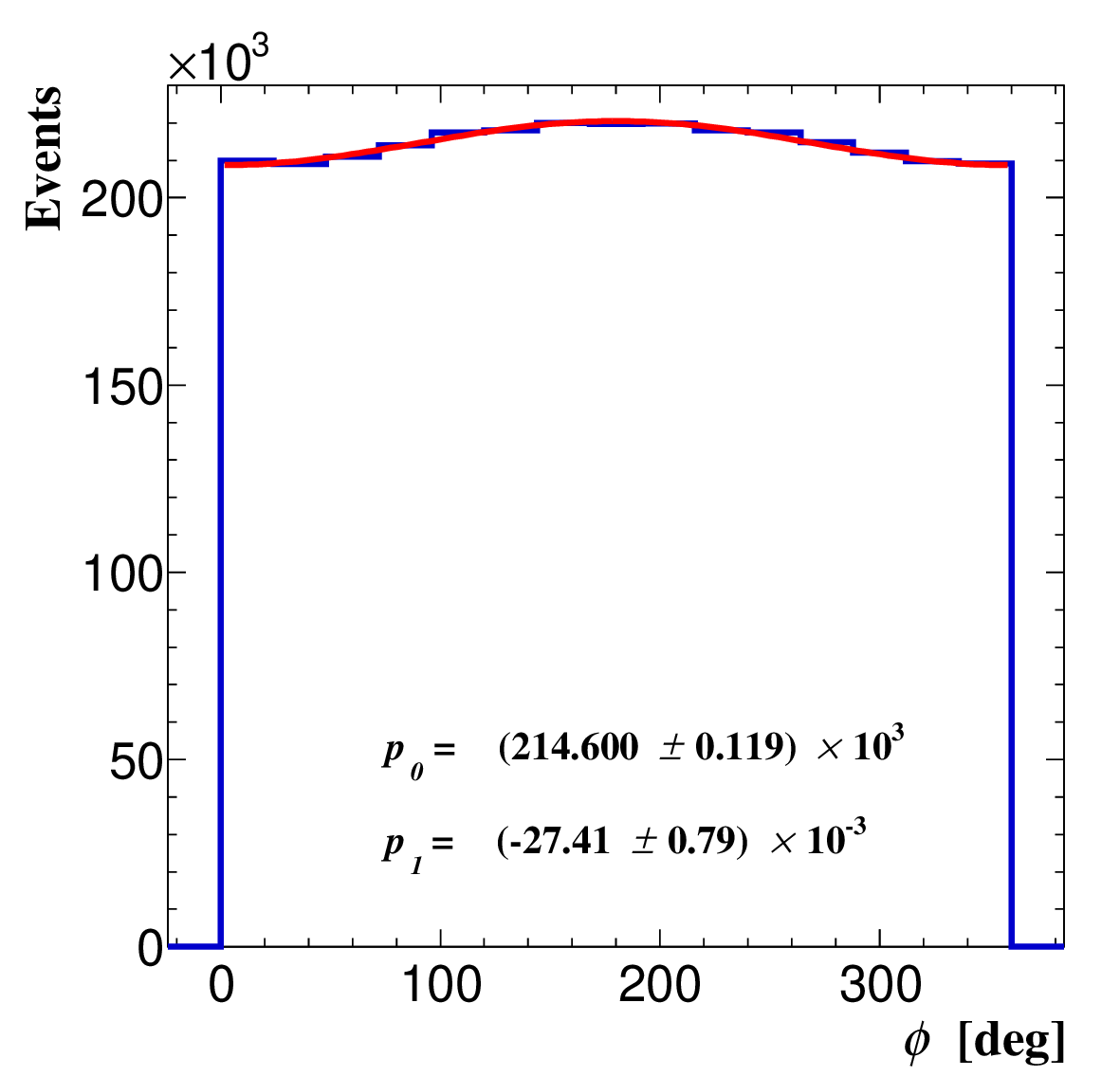}%
    \includegraphics[width=0.445\columnwidth]{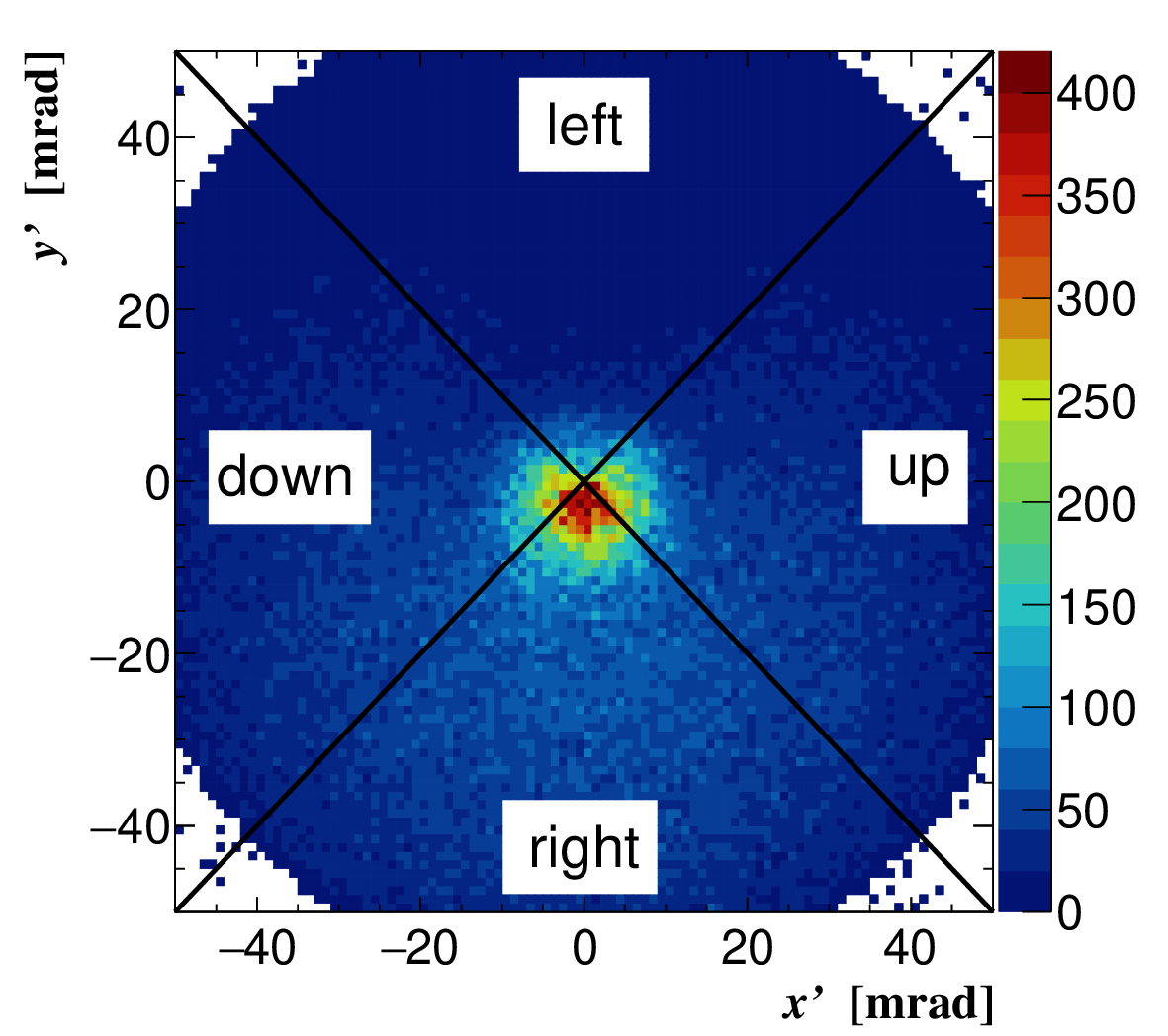}%
}}
\caption{
{\bf (left)} Exemplary distribution of the $\phi$ angle for antiproton tracks. The superimposed red curve denotes the fit according to equation~\ref{eq:fit_funtction}.
{\bf (right)} The event sample was categorized based on the orientation of the vector normal to the scattering plane.
For example: events scattered to the left point upwards and the scattering polar angle $\phi=90^o$. The polarization of the sample in this simulation was equal to 100\%. For better visualization, the analyzing power was set to $A_y = -50\%$ for all events. A maximum scattering angle of 50~mrad was used, since the analyzing power is significantly smaller than 1\% for larger scattering angles.
}
\label{fig:phi_fit} 
\end{figure*}

\begin{figure}
\centering
\begin{minipage}{0.50\textwidth}
    \centering
    \resizebox{0.80\textwidth}{!}{%
       \includegraphics[width=0.415\columnwidth]{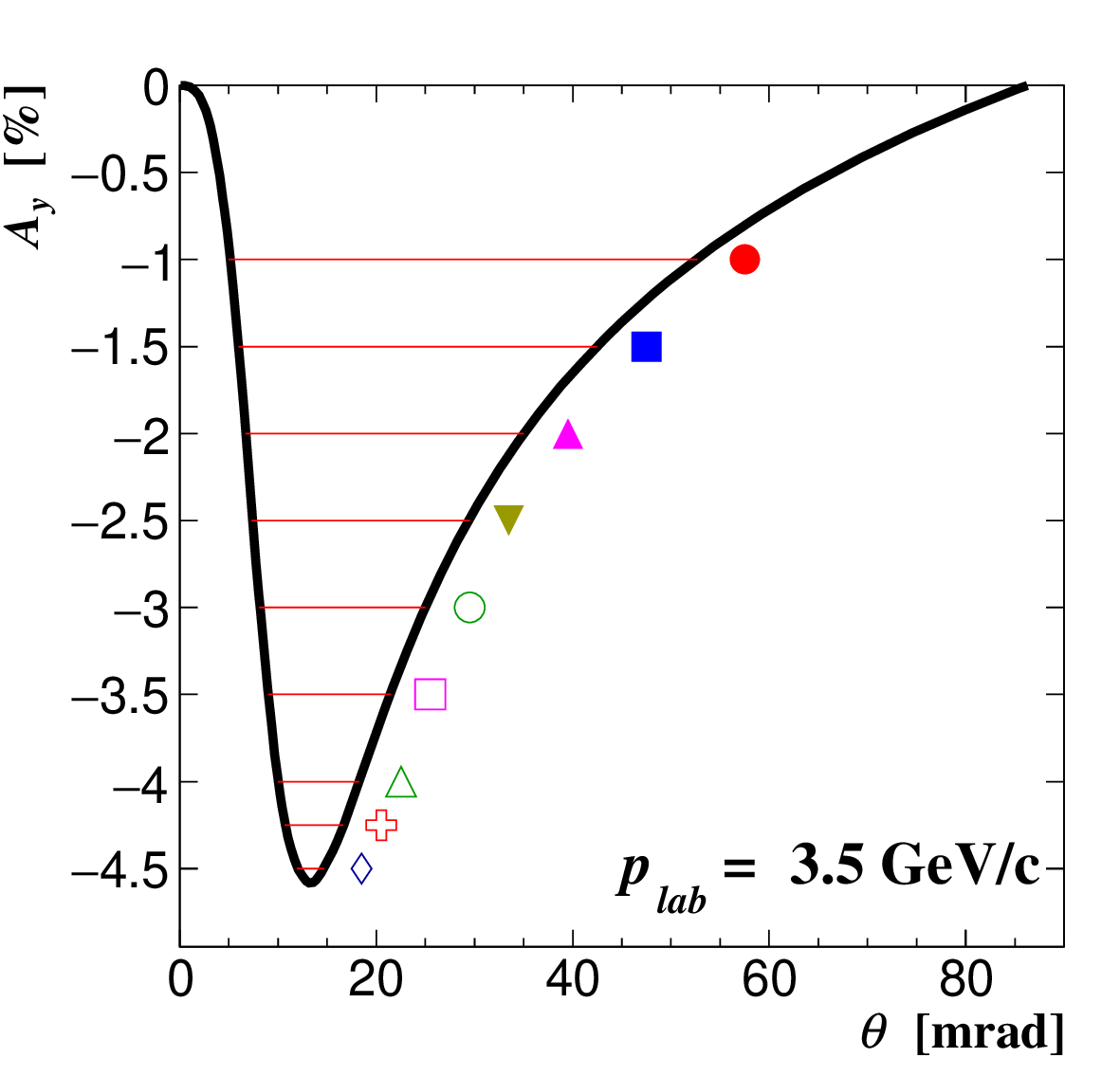}
    }
\end{minipage}\hfill
\caption{
Analyzing power as a function of the scattering polar angle $\theta$ according to predictions given in~\cite{Haidenbauer-pc} for beam momenta $p_{lab}$ = 3.5~GeV/c. 
The red lines denotes the selected  ranges of $A_y$ and the corresponding $\theta$ ranges. Superimposed markers are the same as in Tab.~\ref{tab:angularrange} and Fig~\ref{fig:fom} (right).
}
\label{fig:ay-t-theta} 
\end{figure}

The prepared simulation sample was used to conduct studies on the achievable precision of the measured asymmetry as a function of the assumed degree of antiproton polarization.
In the analysis process, only simulated events containing information about the direction of the antiproton track before and after elastic scattering in the LH$_2$ target were taken into account.
To calculate the $\phi$ angle and polarization for each event, the spin quantization axis was defined along the $y$-axis of the laboratory frame. The resulting simulated $\phi$ distributions were fitted using the function
\begin{equation}
    f(\phi) = p_0 (1+ p_1\cos\phi),
\label{eq:fit_funtction}
\end{equation}
where $p_0$ denotes the normalization factor and $p_1$ reads $<A_yP>$. An example of the $\phi$ angular distribution, with a superimposed fit, is shown in Fig.~\ref{fig:phi_fit}~(left). In addition, to calculate the asymmetry, all events were assigned to one of the following groups: events scattered to the left, right, up, and down. The division is made based on the direction of the vector normal to the scattering plane, and the results are shown in Fig.~\ref{fig:phi_fit}~(right). The asymmetry is calculated based on the number of events in each group, with the statistical uncertainty given by:
\begin{equation}
\sigma(\epsilon) = \frac{1}{\sqrt{N_L + N_R}},
\label{asymmetryError}
\end{equation}
which corresponds to the case of small asymmetries, i.e., \(N_L \simeq N_R\).
Moreover, to identify the optimal range for polarization measurement, it is necessary to determine the scattering angular range that maximizes the mean analyzing power $A_y$. In order to do so, polar angular ranges have been selected to cover different minimal values of the analyzing power, according to the calculations performed by Haidenbauer~\cite{Haidenbauer-pc}. The corresponding angular ranges chosen for the minimal value of the analyzing power are shown in Fig.~\ref{fig:ay-t-theta} and in Table~\ref{tab:angularrange}.
\begin{table}[]
    \centering
   \begin{tabular}{| c | c | c |}
        \hline
        \textbf{\small $A_y$} & \textbf{\small $\theta$ [mrad]} & \textbf{Marker} \\
        \hline
        $A_y \leq$ -1.00\% & 5.00 - 52.70       & \markerCircleFull \\
        \hline
        $A_y \leq$ -1.50\% & 6.00 - 42.50       & \markerBoxFull \\
        \hline
        $A_y \leq$ -2.00\% & 6.70 - 35.00       & \markerTriangleFull \\
        \hline
        $A_y \leq$ -2.50\% & 7.30 - 29.50       &\markerTriangledown \\
        \hline
        $A_y \leq$ -3.00\% & 8.10 - 25.00       & \markerCircleOpen \\
        \hline
        $A_y \leq$ -3.50\% & 9.00 - 21.50       &\markerBoxOpen \\
        \hline
        $A_y \leq$ -4.00\% & 10.00 - 18.23      & \markerTriangleOpen \\
        \hline
        $A_y \leq$ -4.25\% & 10.66 - 16.65      & \squareCross \\
        \hline
        $A_y \leq$ -4.50\% & 11.90 - 14.66      & \markerDiamondOpen \\
        \hline
    \end{tabular}
    \caption{Minimum values of the analyzing power $A_y$ and the corresponding scattering angle $\theta$ range. 
The right column presents the markers used to represent the results for the corresponding scattering angle range in Fig.~\ref{fig:fom}.}
    \label{tab:angularrange}
\end{table}
Such a prepared event sample is then used to calculate the number of events needed to obtain a desired asymmetry and to extract the sensitivity for measuring the possible polarization. 

\section{Results}
\label{sec:results}
\begin{figure*}
\centering{
\resizebox{0.80\textwidth}{!}{%
    \includegraphics[width=0.415\columnwidth]{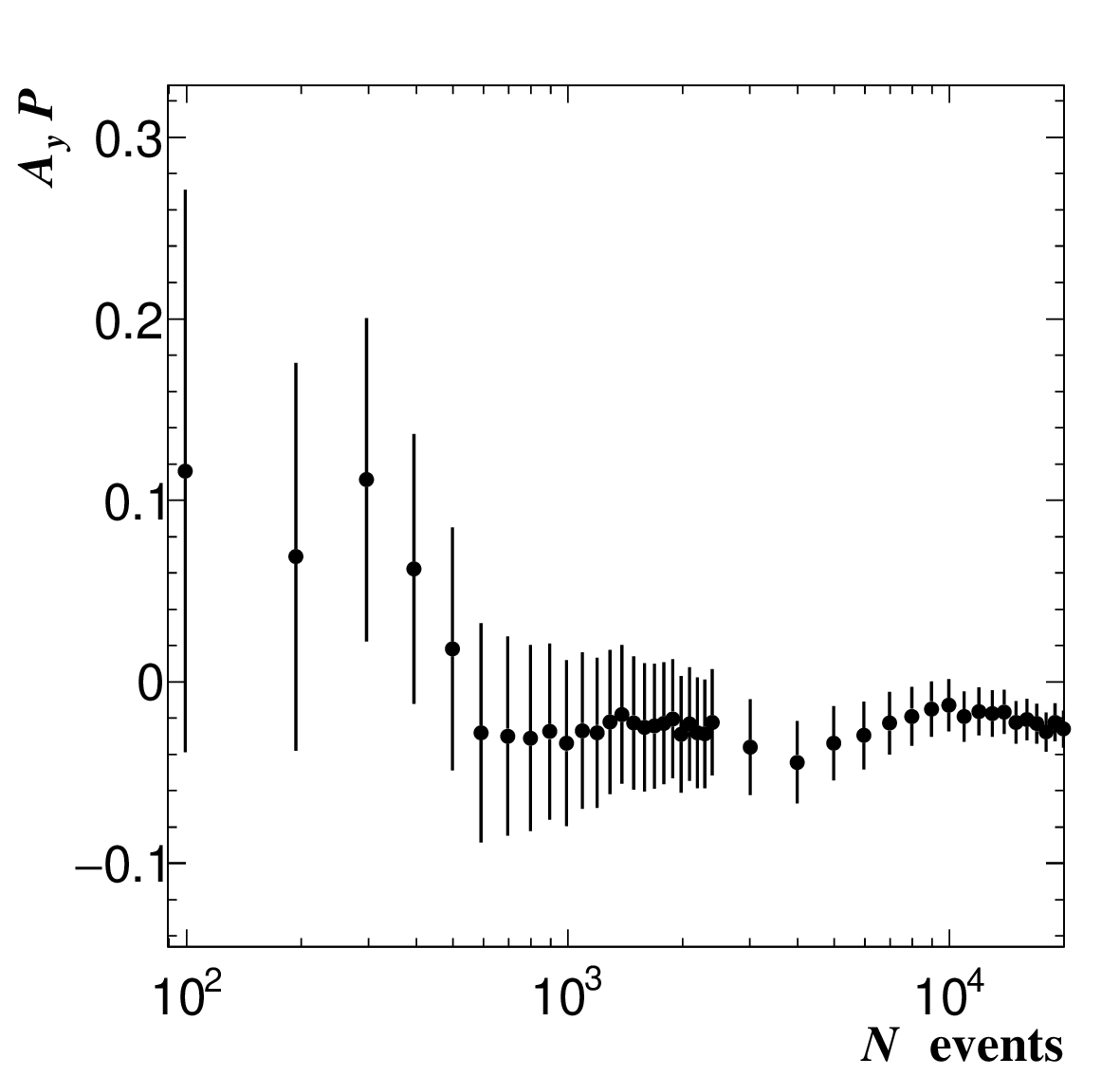}%
    \includegraphics[width=0.415\columnwidth]{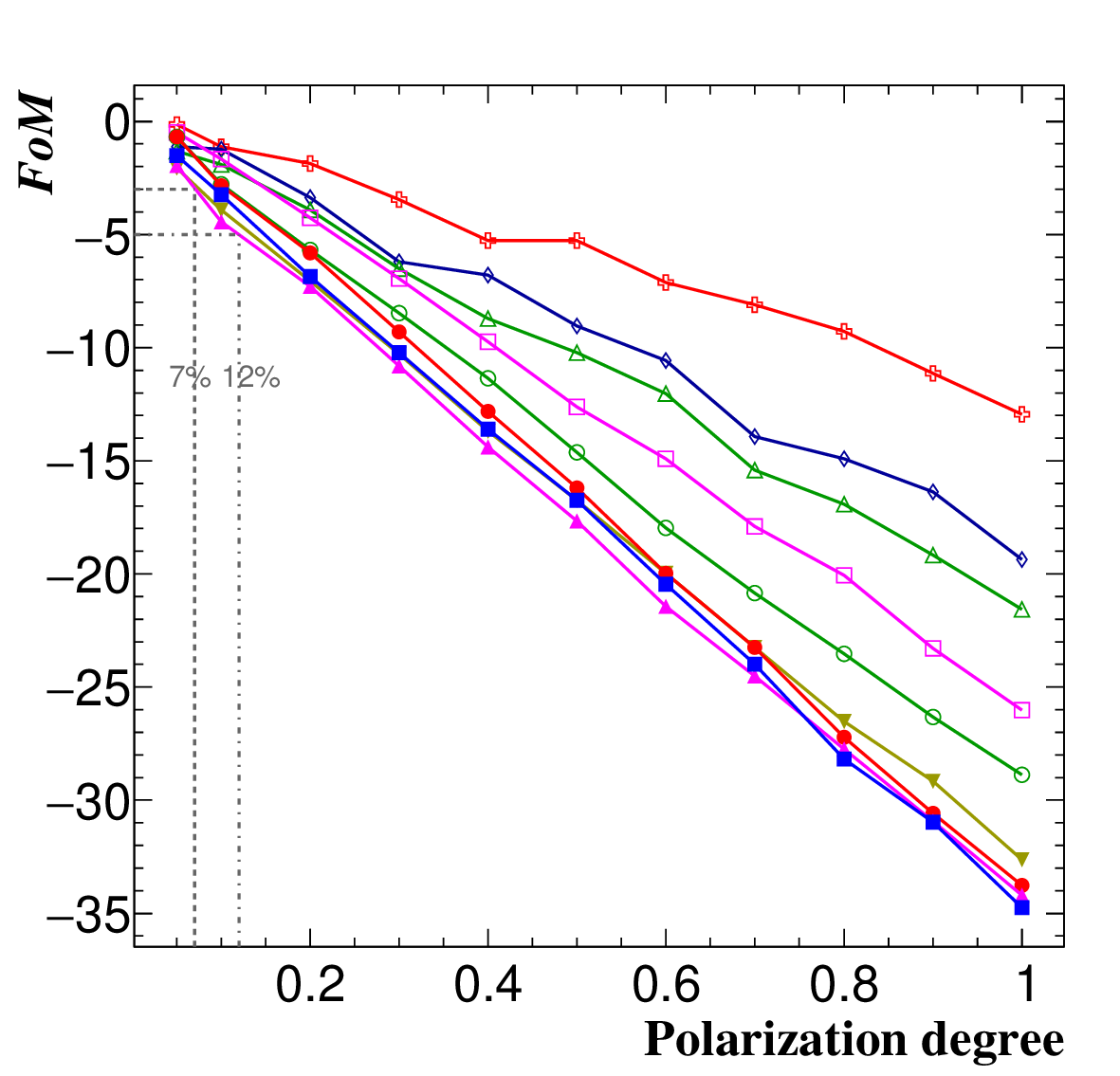}%
}}
\caption{{\bf (left)} Example of the reconstructed $A_y P$ value obtained from the fit according to Eq.~\ref{eq:fit_funtction} as a function of the number of events $N$ with reconstructed primary and scattered tracks in the polar angular range $6<\theta<42$~mrad. The horizontal axis is restricted to $N$ = 20\,000 events to illustrate individual $A_y P$ values together with their statistical uncertainties. {\bf (right)} Figure of Merit ($FoM$) resulting from  the asymmetry $\epsilon$ as a function of the
set polarization for the entire Monte Carlo event sample. Markers indicate the different scattering angular ranges, according to the Tab.~\ref{tab:angularrange}. The gray dashed lines indicates the 7\% and 12\% polarization values, respectively. }
\label{fig:fom} 
\end{figure*}
The presented studies were conducted to investigate the feasibility of determining antiproton polarization, if it exists, in the production process. The cross section for elastic $\bar{p}p$ scattering at the simulated beam momentum, i.e. 3.5~GeV/c is about 1.35~mb in the range -0.002~(GeV/c)\(^2\)~$< t <$~-0.007~(GeV/c)\(^2\)~\cite{E760:1996mur}, which is considered an optimal range for a polarization analysis with a relatively high value of $A_y$. To study the evolution of the results with increasing statistics, the fitting procedure according to Eq.~\ref{eq:fit_funtction} is applied to samples of different sizes. The number of analyzed events is increased in steps of 1000, except for up to 2500 events, where steps of 100 are used. The reconstructed $A_y P$ for each step (increasing event number) is presented in Fig.~\ref{fig:fom}~(left).
The obtained simulation sample shows that the expected asymmetry improves with an increasing number of registered events.
This is evident in the case of an assumed 100\% polarization, as presented in Fig.~\ref{fig:fom}~(left), for the scattering angular range $\theta$ from 6 to 42~mrad in the laboratory frame. Therefore, a high-statistics event sample must be collected during the experiment to achieve reasonable precision for the polarization measurement. To investigate the significance, we have generated $1.6\times 10^6$ scattering events, which were used for the analysis procedure.

For the chosen values of polarization and angular range used in the analysis, 
the figure of merit was extracted, which is defined by $FoM = \frac{A_yP}{\sigma}$, where $\sigma$ denotes the uncertainty in determining the asymmetry, given by Eq.~\ref{asymmetryError}. The results are presented in Fig.~\ref{fig:fom}~(right).
The obtained results show that within the angular region 6.7 $< \theta <$ 35 mrad, a precision of $5\sigma$ is achieved if the polarization is 12\%, and a precision of 3$\sigma$ is reached if the polarization is 7\%.
For any polarization, the achievable precision can be extracted from the analysis. Furthermore, the optimal scattering angular range for the polarization analysis can be determined for each assumed polarization value by examining the $FoM$ as a function of the scattering angle or the peak analyzing power. The analysis shows that the range between 6.7 and 35~mrad yields the most favorable results.

\subsection{Systematic uncertainties}
The present feasibility study is based on Monte Carlo simulations and does not yet allow for a fully quantitative evaluation of systematic uncertainties. In this section, we identify the dominant sources of potential bias and outline the corresponding control strategies. In the proposed measurement, polarization is extracted from a left-right asymmetry $\varepsilon$, given by Eq.~(\ref{eq:asymmetry}), which is related to the antiproton polarization $P$ through the analyzing power $A_y$. Thus, the main systematic uncertainty contributions of the measured $\varepsilon$ are expected to originate from (i) acceptance and efficiency differences between left and right scattering, (ii) antiproton identification purity and background contamination, (iii) the analyzing power input used to relate the measured asymmetry to polarization, and (iv) target and detector alignment.

Differences in detector acceptance and reconstruction efficiency can bias the measured left-right asymmetry if not properly controlled. In the proposed studies, the asymmetry is measured in a detector configuration with a symmetric azimuthal acceptance. Possible effects arising, for example, from local inefficiencies or azimuthally dependent acceptance variations are expected to generate residual asymmetries. These effects can be constrained using data from simultaneously measured pion production channels, for which no transverse polarization signal is expected.

The purity of antiproton identification constitutes another important systematic contribution. An unpolarized background from misidentified particles contributes symmetrically to the event sample and, therefore, reduces the extracted left-right asymmetry. For an antiproton sample purity of approximately $95\%$, the resulting systematic uncertainty on the extracted polarization is expected to be at the percent level or smaller.

Further systematic uncertainty is associated with the analyzing power $A_y$ used to convert the measured asymmetry into polarization. Its uncertainty introduces an overall scale uncertainty on $P$, typically at the level of a few percent, depending on the available input data. However, this scale uncertainty does not affect the sensitivity to a non-zero polarization signal but only the absolute normalization of the extracted polarization.

Misalignment of the target or detector elements may introduce additional left-right asymmetries. However, such effects can be monitored and corrected using reconstructed vertex distributions and symmetry checks.

In general, any systematic uncertainty results in a reduction of the achievable sensitivity for the determination of the polarization which can be compensated by an increase in the statistics. Furthermore, systematic uncertainty can be very well controlled by monitoring simultaneously measured pion asymmetries, as well as reference measurements with positive particles (protons and pions), without a target and at different momenta. For a quantitative estimate of the systematic uncertainties, the
production and analysis of the data are required. 

\section{Conclusions and outlook}
\label{sec:conclusions}
The preparation of a sufficiently polarized antiproton beam for future experiments remains a challenge. A promising approach could be developed if the antiproton production process itself generates a substantial level of polarization.  
The feasibility studies presented in this paper indicate that, for an expected cross section of 
$1.35\,\text{mb}$ and a predicted yield of $1.6\times10^{6}$ events, an integrated luminosity of $1.18\,\text{nb}^{-1}$ (corresponding to about 8 weeks of data taking) would provide sufficient statistics to measure a reasonable antiproton polarization. Assuming nearly $100\%$ detection efficiency, this statistic would allow a 12\% polarization to be differentiated relative to the null hypothesis with a 5$\sigma$ significance and \(7\%\) polarization with a precision of \(3\sigma\). This range of values is particularly relevant for exploring the potential of the antiproton production process as an initial step in preparing a polarized beam. Obtained results clearly demonstrate the feasibility of such a measurement, which is now proposed to be realized in the P371 experiment at the T11 beam line at CERN. 

In conclusion, this work demonstrates that a measurement of the transverse polarization of antiprotons produced in $p+A$ interactions at a few~GeV$/c$ is experimentally feasible within a realistic beam time at the CERN PS. Such a measurement would provide the first direct experimental information on transverse spin effects in antiproton production and would constitute a new observable sensitive to spin dependent dynamics of the $\bar{p}N$ interaction. A verified non-zero transverse polarization of produced antiprotons could  represent a useful input for future experimental developments. In particular, it may contribute to the design of experiments requiring controlled antiproton polarization without relying on additional polarization manipulation techniques. The results presented here establish a well defined experimental basis for future studies of antiproton spin phenomena at intermediate energies.

\bibliographystyle{unsrt}
\bibliography{bibp349}

\end{document}